\documentclass[runningheads,a4paper]{llncs}
\usepackage{subfig}
\usepackage{graphicx}
\usepackage{amssymb}
\usepackage{ctable}
\usepackage{multirow}
\usepackage{dsfont}
\usepackage{tabularx}
\usepackage{bbm}
\usepackage{booktabs}
\usepackage{siunitx}
\usepackage[misc]{ifsym}

\begin{document}

\title{Automatic Myocardial Infarction Evaluation from Delayed-Enhancement Cardiac MRI using Deep Convolutional Networks \vspace{-0.4cm}}
\author{Kibrom Berihu Girum\inst{1,2 (\textrm{\Letter})} \and
Youssef Skandarani\inst{1} \and
Raabid Hussain\inst{1} \and
Alexis Bozorg Grayeli\inst{1,4} \and
Gilles Cr\'ehange\inst{1,2,3} \and
Alain Lalande\inst{1}}

\institute{ImViA Laboratory, University of Burgundy, Dijon, France \\
 \email{kibrom-berihu\_girum@etu.u-bourgogne.fr} \and Radiation Oncology Department, CGFL, Dijon, France \\
  \and
 Radiation Oncology Department, Institut Curie, Paris, France \\
  \and ENT Department, CHU Dijon, Dijon, France \\
 }

\titlerunning{Myocardial Infarction Evaluation from DE-MRI and Clinical Information}  
\authorrunning{K.B. Girum et al.}
\maketitle             

\begin{abstract}
In this paper, we propose a new deep learning framework for an automatic myocardial infarction evaluation from clinical information and delayed enhancement-MRI (DE-MRI). The proposed framework addresses two tasks. The first task is automatic detection of myocardial contours, the infarcted area, the no-reflow area,  and the left ventricular cavity from a short-axis DE-MRI series. It employs two segmentation neural networks. The first network is used to segment the anatomical structures such as the myocardium and left ventricular cavity. The second network is used to segment the pathological areas such as myocardial infarction, myocardial no-reflow, and normal myocardial region. The segmented myocardium region from the first network is further used to refine the second network's pathological segmentation results. The second task is to automatically classify a given case into normal or pathological from clinical information with or without DE-MRI. A cascaded support vector machine (SVM) is employed to classify a given case from its associated clinical information. The segmented pathological areas from DE-MRI are also used for the classification task. We evaluated our method on the 2020 EMIDEC MICCAI challenge dataset. It yielded an average Dice index of 0.93 and 0.84, respectively, for the left ventricular cavity and the myocardium. The classification from using only clinical information yielded 80\% accuracy over five-fold cross-validation. Using the DE-MRI, our method can classify the cases with 93.3\% accuracy. These experimental results reveal that the proposed method can automatically evaluate the myocardial infarction.

\keywords{Heart . Cardiac MRI segmentation . CNNs . Myocardial infarction . Myocardial no-reflow .  Classification . Clinical information.}
\end{abstract}
\section{Introduction}
Cardiovascular diseases (CVDs) are the leading cause of death in the world \cite{mozaffarian2017global}. Among the common CVDs,  myocardial infarction (MI) is a specific cardiovascular disease. The state of the heart after myocardial infarction receiving revascularization requires careful evaluation of the myocardial segment's functionality. It requires assessment of myocardial infarction and myocardial no-reflow (persistent microvascular obstruction area) regions using magnetic resonance imaging (MRI). This is often done using a delayed enhancement-MRI (DE-MRI), which involves an MRI examination following several minutes of a contrast agent injection. Accurate and automatic patient classification is profoundly essential in this scenario. Moreover, accurate and automatic segmentation of the clinically essential regions such as myocardial infarction and myocardial no-reflow can help to assess tissue functionality and thereby make more accurate treatments based on the severity of the disease. Developing an accurate and automatic cardiac image segmentation framework is a challenging task due to low contrast boundary regions, different sizes and shapes of targets, and respiratory motion. Indeed, variations in image acquisition settings in different clinical setups are other challenges in developing accurate and robust medical image segmentation methods \cite{girum2020fast}.

Recently, deep learning approaches using convolutional neural networks have provided the state of the art results in medical image analysis tasks. They have successfully been applied in classification and segmentation tasks. In this paper, we propose a deep learning framework for automatic and accurate evaluation of myocardial infarction from clinical information with and without the delayed-enhancement cardiac-MRI. The proposed framework has two main applications. The first application is to automatically detect the myocardial contours, the infarcted area, the permanent microvascular obstruction area (no-reflow area), and segments the left ventricular cavity from a short-axis DE-MRI. The second application is to automatically classify a given case into normal or pathological using the provided clinical information with or without DE-MRI. The provided clinical information includes sex, age, overweight, arterial hypertension, diabetes, familial history of coronary artery disease, Troponin, Killip Max, ejection fraction, ventricular natriuretic peptide, and the ST segment.

We evaluated our method on a MICCAI 2020 challenge dataset, named automatic Evaluation of Myocardial Infarction from the Delayed-Enhancement Cardiac MRI (EMIDEC) \cite{lalande@2029EMIDEC}. The experimental results show that the proposed method can be used to automatically evaluate the myocardial infarction from cardiac DE-MRI as well as from clinical information. Moreover, the proposed method can accurately segment the anatomical structures such as the myocardium and the left ventricular cavity.  

\section{Proposed framework}
\subsubsection{Segmentation} 
\begin{figure}[ht]
	\centering
	\includegraphics[width=1\linewidth]{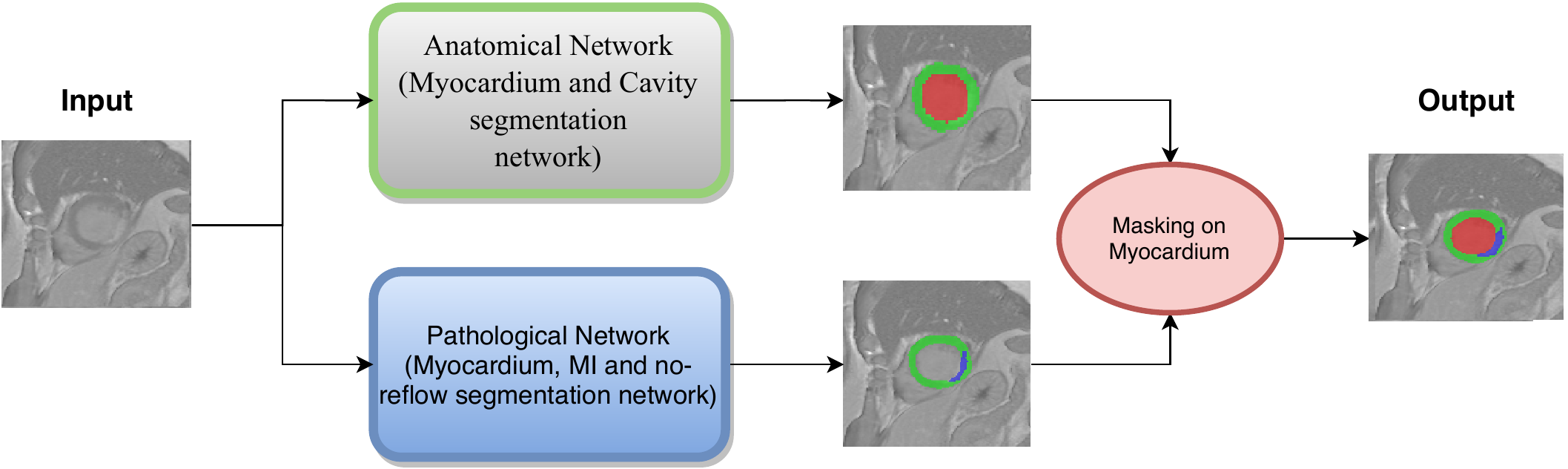}
	\caption{The schematic representation of the proposed segmentation framework. The anatomical network is used to segment the left ventricular cavity and the myocardium. The pathological network is used to detect the pathological areas (myocardium infarction (MI) and no-reflow) and the healthy or normal myocardium regions. The red, green, and blue colors, respectively, show the left ventricular cavity, the myocardium,  and the myocardium infarction. }
	\label{fig_model}
\end{figure}

To segment the anatomical and pathological (if present) structures from DE-MRI, we proposed a new deep learning framework (Fig. \ref{fig_model}). It employed two encoder-decoder style convolutional neural networks \cite{ronneberger2015u}. The first network, the anatomical network, was used to segment the anatomical structures such as the myocardium and the left ventricular cavity. The second network, the pathological network, was used to segment the areas such as the myocardium infarction, the no-reflow, and the normal or healthy myocardium regions.  The segmentation results from the pathological network are then masked by the segmented myocardium region from the anatomical network. It enabled us to constrain the segmentation results of the pathological network to be inside the myocardium area. However, note that the anatomical and pathological networks were trained separately. The segmented myocardium area from the anatomical network was the final myocardium segmentation. The predicted segmentation from the anatomical structure segmentation network and the masked pathological areas from the pathological network are then merged to yield the final four label segmentation output. 

The encoder-decoder networks' building block consists of a repeated application of 2D $3$$x$$3$ convolution, exponential linear unit (ELU) activation function, batch normalization followed by squeeze-and-excitation network \cite{hu2018squeeze}, and a $2$$x$$2$ max pooling operation with stride 2 for down-sampling. Starting with 32 feature maps, after each max-pooling layer the feature channels are then doubled. The up-sampling part consists of $2$$x$$2$ up-convolution and concatenation layer, followed by repeated 2D $3$$x$$3$ convolution, exponential linear unit (ELU) activation function, batch normalization, and squeeze-and-excitation network \cite{hu2018squeeze}. The output 2D convolutional layer activation function was softmax for both pathological and anatomical networks. Each network was trained by calculating the average of Dice and cross-entropy loss.

\subsubsection{Classification}
\begin{figure}[ht]
	\centering
	\includegraphics[width=0.9\linewidth]{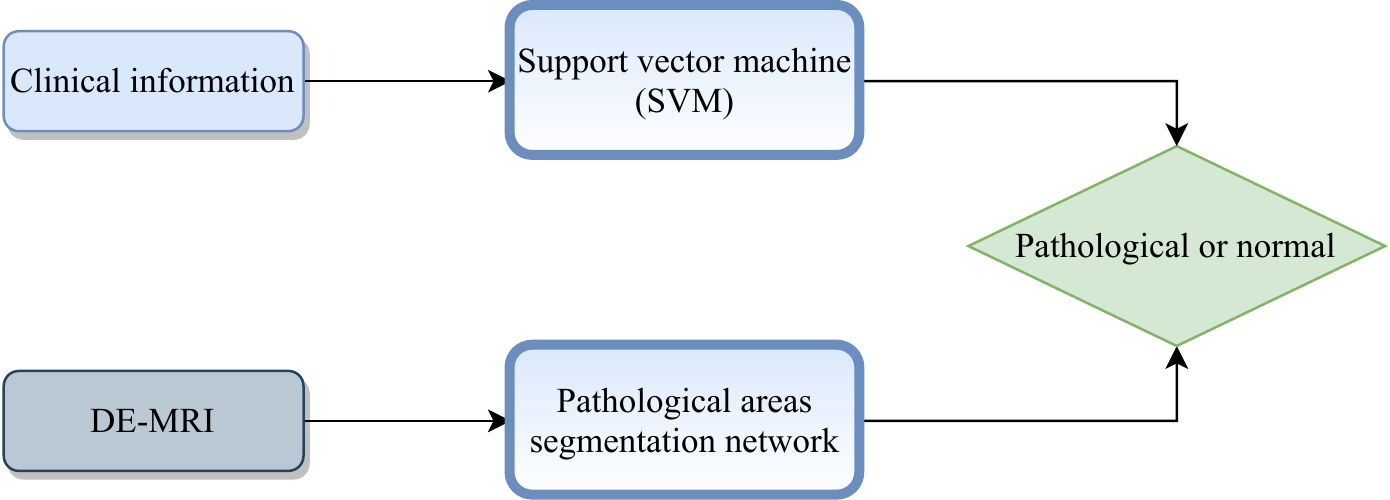}
	\caption{The schematic representation of the proposed classification framework using the provided clinical information and cardiac DE-MRI.}
	\label{fig_model_classificaiton}
\end{figure}

The classification task revolves around the clinical information for each patient. The proposed approach used a Support Vector Machine (SVM) with a linear kernel and a hinge loss \cite{Bishop07} as an initial feature selection step. The selected features were then passed into another SVM with a radial basis function kernel to refine the separation boundary between the two classes (i.e., normal or pathological). The cascaded SVM was designed to remove redundant features at the first step and provide robust features for final classification in the second step. Moreover, classification using the DE-MRI was performed from the segmentation results by checking if the segmented areas consist of either the myocardium infarction or the no-reflow. If a given case consists of pathological areas in two or more slices, it was considered pathological otherwise normal. The classification pipeline is shown in Fig. \ref{fig_model_classificaiton}.

\subsubsection{Implementation details}
The proposed framework was implemented on an i7 computer with 32 GB RAM and a dedicated GPU (NVIDIA TITAN Xp, 12 GB). Python, Keras API with Tensorflow backend, was used to implement the method. We trained the model using the ADAM optimizer for $500$ epochs with a learning rate of $10^{-3}$ and early stopping criteria of 200 \cite{kingma2014adam}. All convolutional neural networks' parameters were initialized using He initialization \cite{he2015delving}. Moreover, other hyper-parameters values, such as the ADAM optimizer parameters, were Keras' default values. 

\section{Experimental setup and results}
\label{sec::setup}
\subsubsection{Dataset and evaluation criteria}
\label{sec::data}
To evaluate the proposed method, we selected the EMIDEC MICCAI 2020 challenge dataset \cite{lalande@2029EMIDEC}. It consisted of 100 training DE-MRI images, of which 67 are with pathological and 33 normal cases. The 50 held out testing data consisted of 33 pathological conditions and 17 normal cases. The training dataset was divided randomly into 85 training and 15 validation cases. The validation cases were composed of 10 pathological and five normal cases. As the provided EMIDEC dataset was with different image resolutions, we resized it to $256x256$ before feeding it to the network. Each given case was also pre-processed by zero-centering the intensity values and normalizing it by the standard deviation. Although the proposed segmentation networks are in 2D, we evaluated the results in 3D after stacking each predicted 2D images per case. Thus, the evaluation was in 3D using the Dice index (DSC), Hausdorff distance (HD), and relative volume difference (RVD) for the segmentation task. For the classification task, we used the accuracy metric. 

\subsubsection{Experimental results}
\label{sec:results}
Experimental results show that the proposed method can segment the left ventricular cavity and the myocardium cardiac structures with an average Dice index of 0.93 and 0.84, and an average 3D HD value of 6.5 mm and 8.9 mm, respectively. Quantitative segmentation results are shown in Table \ref{table_results}. The proposed SVM-based classification method yielded a classification accuracy of 80\%  using only the clinical information in 5-fold cross-validation. Moreover, our method based on the segmented pathological areas from the pathological network classifies with 93.3\% accuracy using DE-MRI only. However, although the proposed method detects the pathological areas, it appeared to under segment them yielding less overlapping area ratio. For example, it yielded an average Dice index of 0.4 and 0.67 and an average relative volume difference of 0.242 and 0.375 for the myocardial infarction and no-reflow, respectively.

\begin{table}[h]
	\caption{Quantitative segmentation results on the EMIDEC dataset (15 validation sets). Values are expressed in mean $\pm$ standard deviation (std), minimum (Min.), and maximum (Max.). } 
	\centering		
	\begin{tabular}{  l  l l l | l l l | l l l}
		 {}  &  \multicolumn{9}{c}{Metric}  \\  \cline{2-10}
		 Target &  \multicolumn{3}{c}{DSC}  & \multicolumn{3}{c}{ HD(in mm)} & \multicolumn{3}{c}{ RVD}\\
		\cline{2-5}
		\cline{6-8}
		\cline{8-10}
		 & Mean $\pm$ std & Min. & Max. &  Mean $\pm$ std & Min. & Max. & Mean $\pm$ std & Min. & Max.\\
		\hline
		 
		 LV &  0.93$\pm$0.018 & 0.90 & 0.96 & 6.5$\pm$3.7 & 3.9 & 19.2 &0.06$\pm$0.05 & 0.003 & 0.20\\
		 
		 MYO &  0.84$\pm$0.03 & 0.75 & 0.88 & 8.9$\pm$5.3 & 4.9 & 25.9 &0.08$\pm$0.08 & 0.004 & 0.27\\
		\hline
	\end{tabular}
	\label{table_results}
\end{table}

The proposed method showed promising performance to segment the anatomical structures such as the myocardium and the left ventricular cavity from DE-MRI. To extend this approach, shape prior based deep learning methods could help constrain the segmentation of the anatomical structures \cite{girum2020deep,oktay2017anatomically,girum2019deep}. Post processing methods using convolutional auto-encoders \cite{larrazabal2020post,painchaud2020cardiac} could also constrain the segmentation process by guaranteeing the correct shape of the anatomical structures. Moreover, the clinical information could be used to constrain the segmentation process of the pathological areas.

\begin{figure}[ht]
	\centering
	\includegraphics[width=1\linewidth]{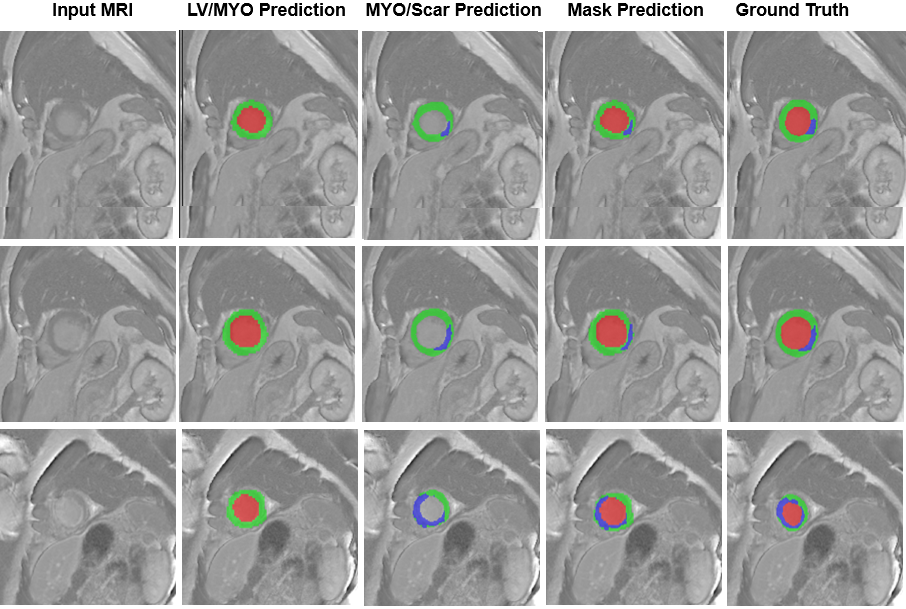}
	\caption{Examples of segmentation results on the EMIDEC validation set. [Column 1] Input DE-MRI; [Column 2] Predicted left ventricular cavity (LV) and myocardium (MYO); [Column 3] Predicted normal MYO areas and myocardium infarction (MI) areas; [Column 4] All predicted structures including the LV, MYO, and MI; [Column 5] Ground truth level from expert. The red, green, and blue colors respectively show the LV, the MYO, and the MI. 
	\label{fig_seg_results}
	}
\end{figure}

\section{Discussion and conclusion}
\label{sec::conclu}
In this paper, we proposed a deep learning framework to automatically segment the left ventricular cavity and the myocardium regions from DE-MRI. We also proposed a deep learning framework to automatically detect pathological areas, particularly the myocardium infarction and the no-reflow areas, from the DE-MRI. Moreover, the classification of a given case into pathological or normal was performed using the provided clinical information as well as using the DE-MRI.

Experimental results, obtained using the EMIDEC MICCAI 2020 challenge dataset, revealed that the proposed framework can accurately segment the heart structures. Moreover, our method can detect pathological areas. Besides, the classification of a given case can be performed using either clinical information or DE-MRI, which could benefit practitioners in assessing the state of the heart. However, the classification accuracy from DE-MRI seems to outperform the approach using only the clinical information. Next, shape prior based methods will be explored to constrain the segmentation process of anatomical structures such as the myocardium and the left ventricular cavity. The clinical information could also be used to constrain the pathological areas' segmentation process.   

\subsection*{Acknowledgment} The authors would like to thank NVIDIA for providing GPU (NVIDIA TITAN Xp, 12 GB) through their GPU grant program.
%
%
{
\bibliographystyle{splncs}
\footnotesize
\bibliography{main}
}
\end{document}